\documentstyle[epsf]{l-aa}

\def\kmsec{\hbox{${\rm km\ s^{-1}}$}}
\newcommand{\aj}{AJ}         
\newcommand{\apj}{ApJ}       
\newcommand{\baas}{BAAS}     
\newcommand{\mnras}{MNRAS}   
\newcommand{\qjras}{QJRAS}   
\newcommand{\stel}{Sky Telesc.}  
\newcommand{\spsrev}{Space Sci.~Rev.} 

\begin{document}

   \thesaurus{08         
              (08.09.2;  
               08.14.2;  
               08.02.1;  
               08.12.1)} 

\title{The Status of Nova Orionis 1667}

\author{F. A. Ringwald\inst{1} \and T. Naylor \inst{2}}

\institute{Department of Astronomy and Astrophysics, The Pennsylvania
State University, 525 Davey Laboratory, \\
University Park, PA 16802-6305, USA (ringwald@astro.psu.edu)\\
\and
Department of Physics, Keele University, Keele, Staffordshire,
ST5~5BG, UK (timn@astro.keele.ac.uk)}

\offprints{F. A. Ringwald}

\date{Received 27.2.97; accepted 9.5.97 }

\maketitle


\begin{abstract}

Images taken in 1.3\arcsec\ seeing show that a claimed hibernating nova,
Candidate 5 shown in a {\it Sky \& Telescope\/} Newswire item, is found to
be a visual binary, with a 1.1\arcsec\ separation between components. 
Spectra reveal the components to be emission-line M0 and K7 stars. 


\end{abstract}

\section{Introduction}

The hypothesis that novae hibernate in the millenia between outbursts
predicts that very old novae should be much less luminous than those seen
in the 20th century, because mass transfer in these systems has diminished
or stopped entirely (Shara et al.~1986). Nova hibernation theory has a
serious drawback, though: not one old nova has ever clearly been shown to
be hibernating (Naylor et al.~1992; Mukai \& Naylor 1995; Somers et
al.~1996, 1997).  To test the theory, it is therefore essential to recover
novae as old as possible.  Among the oldest novae listed in the catalog of
Duerbeck (1987) is V529 Ori, or Nova Ori 1667. 

Our knowledge of V529 Ori has had a complex history.  It was discovered at
$m_{vis} = 6$ by J.~Hevelius during a lunar occultation.  Although claimed
several times to be a recurrent nova, it probably is not one: all but the
first account have been discredited as duplications of the original
observation (Ashbrook 1963; Duerbeck 1987; Webbink et al.~1987).  Details
surrounding the original observation, such as the date of observation,
have also been called into question (Ashworth 1981): it should be 1678,
not 1667. 

Recently, a Newswire item in {\it Sky \& Telescope\/} magazine reported
that this nova had been recovered, and was in deep hibernation. This
Newswire item was originally uncredited, but it was written by R. T.
Fienberg (R.  T. Fienberg, 1996, private communication) and so will be
referenced here as Fienberg (1995). It was based on a poster paper by
Wagner et al.~(1994), although Nova Ori 1667 is not mentioned in the
proceedings abstract.

\section{Imaging}


We took three $I$-band images and one H$\alpha$ image of the field of V529
Ori on 1993 August 15 and 16 UT with the 1.0-m Jacobus Kapteyn Telescope
at La Palma. All images were taken in in 1.3\arcsec\ seeing. An EEV CCD
was used at the f/15 Cassegrain focus, which yielded a pixel scale of
0.31\arcsec\ pixel$^{-1}.$ A median of three bias frames was subtracted
from each image, and the data were then flat-fielded using images of the
twilight sky.  We combined the $I$-band exposures into one image
representing a total exposure of 1800 s. The H$\alpha$ image was a 1000-s
exposure. 

The resulting images are shown in Fig.~1. Star 5, using the numbers in the
image of Fienberg (1995), was reported by this author to be the
hibernating nova. Our images clearly show there are two components. 
Astrometry with the Guide Star Catalog (Russell et al.~1990) showed the
1950.0 positions, to within 0.4\arcsec\ absolute and 0.1\arcsec\ relative,
of component A to be $\alpha$ 05~55~15.574~~$\delta$~+20~15~55.88, and of
component B to be $\alpha$ 05~55~15.638~~$\delta$~+20~15~55.22.

\section{Spectra} 


Spectra of both components were taken on 1995 February 8, beginning at
1:56 and 2:29 UT, with the ISIS spectrograph of the 4.2-m William Herschel
Telescope at La Palma. The R158R and R158B gratings were used with TEK
CCDs. The slit was 0.73\arcsec\ wide on the sky, and was set to a position
angle of 26.0 degrees, perpendicular to the line intersecting the centers
of both stars. The parallactic angle at the time varied from 68.5 to 67.5
degrees. Seeing was 0.9 -- 1.0\arcsec, and weather was photometric. The
dispersion was 2.9 \AA\ pixel$^{-1},$ and the exposure times for both sets
of red and blue exposures was 1800 s. A standard reduction was carried
out, in which the frames were debiased, flat-fielded with exposures of
tungsten lamps, wavelength-calibrated using Cu-Ne and Cu-Ar lamps,
sky-subtracted, and flux-calibrated with the spectrum of the standard
Feige 34 (Stone 1977). The spectrum of a nearby F8 star, SAO 077760, was
also taken, to map the telluric absorption bands to allow their removal. 

The resulting spectra of both components of Candidate 5 are shown in
Fig.~2. The signal-to-noise ratios are about 20 near H$\alpha$ in both red
spectra and about 5 for star A and 10 for star B near H$\beta$ in the blue
spectra. The only features obvious in the red spectra are H$\alpha,$ in
emission, and the TiO absorption bands at $\lambda\lambda$\,7165 and 7665
\AA. The blue spectra show only faint Mg b $\lambda$ 5175 \AA\ in star B. 
By comparison with Vilnius dwarf spectra (Sviderskiene 1988), we estimate
spectral types of M0.5 $\pm$ 0.5 for star A and K7.5 $\pm$ 0.5 for star B. 
H$\alpha$ has equivalent widths of 7.3 $\pm$ 0.5 \AA\ and 22.8 $\pm$ 0.8
\AA\ and full-widths at half-maxima of $290 \pm 15$~\kmsec\ and $400 \pm
15$~\kmsec\ in stars A and B, respectively.  Since radial velocity
standards were not taken, it is impossible to quote reliable absolute
values for these stars' radial velocities, but the H$\alpha$ lines show a
difference of $46 \pm 9$~\kmsec\ between stars A and B. 

Despite probable light losses through the slit, we deconvolved
Johnson-Kron broad-band magnitudes from the spectra. The only bands
completely covered were $B$ and $I.$ We found $I = 18.1 \pm 0.1$ for star
A and $I = 18.5 \pm 0.1$ and $B - I = 4.5 \pm 0.2$ for star B; low
signal-to-noise in the blue prevented such a measurement for A.  Assuming
that star B is a K7 -- M0 dwarf, its colors (Bessell 1991) and the
extinction law of Howarth (1983) would imply $E(B - V) = 0.32 \pm 0.05$
and a distance between 1.5 and 1.9 kpc.  Although star A is brighter, it
is redder, although these spectra cannot distinguish its luminosity class.

\section{Discussion} 

The component stars of most classical novae are K -- M dwarfs orbiting
white dwarfs.  Warner (1995, equation 9.54) uses the calculations of
Prialnik (1986) to find that $T_{eff} = 4.8 \times 10^5
[t/0.1]^{-0.28}~\rm{K},$ where $T_{eff}$ is the effective temperature and
$t$ is the time in years since eruption.  This applies to a nova with a
1.25-$M_{\odot}$ white dwarf, and novae with massive white dwarfs should
have much higher discovery probabilities (Ritter et al.~1990). The white
dwarf in a 300-year-old nova should therefore have $T_{eff} \approx
50,000$~K.  White dwarfs this hot are observed to have $M_B = 8.8$
(Wesemael, Green, \& Liebert 1985) and $B - V = -0.3$ (Sion \& Liebert
1977).  In $B$ and $V,$ therefore, any white dwarf present should have
absolute magnitudes comparable to or brighter than those of M0 or K7
dwarfs, with $M_B = 10.27$ and 9.55 and $M_V = 8.86$ and 8.23 (Bessell
1991), respectively.  We see no obvious sign of a hot white dwarf in our
red or blue spectra, however, nor of an accretion disk: if either
Candidate 5a or 5b is the hibernating nova, it is in very deep hibernation
indeed.  Alternatively, these could just be M0 and K7 stars, and the nova
has not yet been recovered.  These stars' H$\alpha$ equivalent widths seem
high for M0 or K7 dwarfs, which more typically are near 2 \AA\ (Linsky et
al.~1982).  However, T Tauri stars abound in Orion, and H$\alpha$ emission
from their accretion disks has equivalent widths ranging from $< 10$ to $>
90$~\AA\ (Jaschek \& Jaschek 1987).  Radial velocity studies could
demonstrate whether either Candidates 5a or 5b are close binary stars.
Until then, however, or until a more convincing candidate has been found,
this nova---if indeed it was a nova---should be considered unrecovered.

\begin{acknowledgements}
F. A. R. thanks PPARC for travel funding. T. N. holds a PPARC Advanced
Fellowship. The Jacobus Kapteyn and William Herschel telescopes are
operated on La Palma by the Royal Greenwich Observatory at the Spanish
Observatorio del Roque de los Muchachos of the Instituto de Astrofisica de
Canarias. Some analysis was done with the ARK software on the Keele
STARLINK node. \end{acknowledgements}

\clearpage  

\begin{figure} 
\epsfysize=150mm
\epsfbox[0 150 468 619]{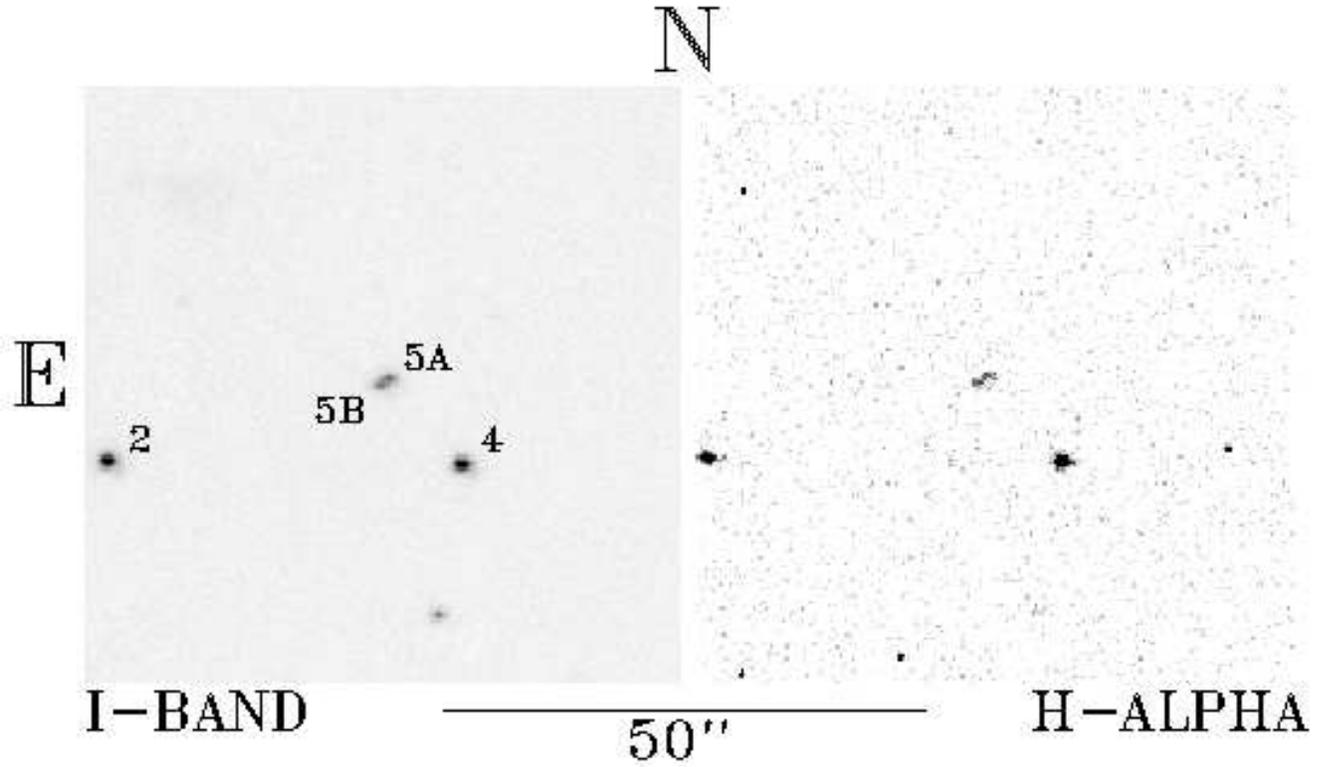}
\caption[ ]{Images in $I$ and H$\alpha$ of the field of Nova Ori 1667. The
stars are labeled with the numbers of Fienberg (1995). The visual binary
in the center of the field is Candidate 5, the claimed hibernating nova.}
\end{figure}

\clearpage

\begin{figure}
\epsfysize=150mm
\epsfbox[0 150 468 619]{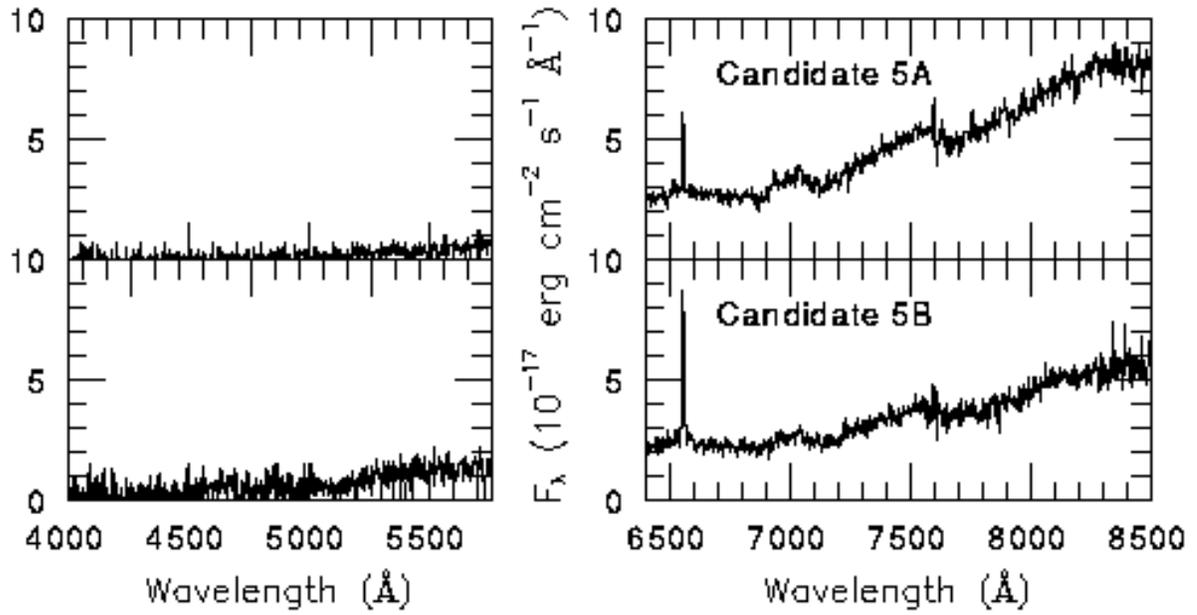}
\caption[ ]{Spectra of components A and B of Candidate 5.  Telluric
absorption bands were mapped and removed from these sky-subtracted
spectra. The TiO $\lambda\lambda$\,7165 and 7665 \AA\ bands are evident in
both red spectra, as is H$\alpha$ in emission.}
\end{figure}

\end{document}